\documentclass[article,preprint,superscriptaddress,preprintnumbers,floatfix, endfloats]{revtex4}

\usepackage[pdftex]{graphicx}
\usepackage{amssymb}
\usepackage{array}
\usepackage{verbatim}
\usepackage[ansinew]{inputenc}
\usepackage[paperwidth=216mm,paperheight=280mm,centering,hmargin=2.54cm,vmargin=2.54cm]{geometry}
\usepackage[english]{babel}
\usepackage{setspace}
\usepackage{subfig}
\usepackage{natbib}
\begin{document}
\DeclareGraphicsExtensions{.pdf,.png,.jpg,.eps,.tiff}
\title{Fabrication and operation of a two-dimensional ion trap lattice on a high-voltage microchip}
\author{R. C. Sterling}
\affiliation{Department of Physics and Astronomy, University of Sussex, Brighton, BN1 9QH, UK}
\author{H. Rattanasonti}
\affiliation{School of Electronics and Computer Science, University of Southampton, Highfield, Southampton, UK, SO17 1BJ}
\author{S. Weidt}
\affiliation{Department of Physics and Astronomy, University of Sussex, Brighton, BN1 9QH, UK}
\author{K. Lake}
\affiliation{Department of Physics and Astronomy, University of Sussex, Brighton, BN1 9QH, UK}
\author{P. Srinivasan}
\affiliation{School of Electronics and Computer Science, University of Southampton, Highfield, Southampton, UK, SO17 1BJ}
\author{S. C. Webster}
\affiliation{Department of Physics and Astronomy, University of Sussex, Brighton, BN1 9QH, UK}
\author{M. Kraft}
\affiliation{School of Electronics and Computer Science, University of Southampton, Highfield, Southampton, UK, SO17 1BJ}
\affiliation{University of Duisburg-Essen, Faculty of Engineering Sciences, Bismarkstrasse 81, Duisburg D-47059, Germany}
\author{W. K. Hensinger\footnote{W.K.Hensinger@Sussex.ac.uk}}
\affiliation{Department of Physics and Astronomy, University of Sussex, Brighton, BN1 9QH, UK}
\email{W.K.Hensinger@sussex.ac.uk}
\begin{abstract}

Microfabricated ion traps are a major advancement towards scalable quantum computing with trapped ions. The development of more versatile ion-trap designs, in which tailored arrays of ions are positioned in two dimensions above a microfabricated surface, would lead to applications in fields as varied as quantum simulation, metrology and atom-ion interactions. Current surface ion traps often have low trap depths and high heating rates, due to the size of the voltages that can be applied to them, limiting the fidelity of quantum gates. Here we report on a fabrication process that allows for the application of very high voltages to microfabricated devices in general and use this advance to fabricate a 2D ion trap lattice on a microchip. Our microfabricated architecture allows for reliable trapping of 2D ion lattices, long ion lifetimes, rudimentary shuttling between lattice sites and the ability to deterministically introduce defects into the ion lattice.

\end{abstract}
\maketitle

The introduction of microfabrication techniques to the field of ion trapping has lead to the development of impressive microfabricated radio-frequency (rf) ion trap devices \cite{IQT2}. Using such devices, many of the building blocks for scalable quantum computing have been demonstrated \cite{09:home}, including ion combination and separation and junction shuttling \cite{racetrack,combi} as well as two-qubit gate operations \cite{uwaveQL, 03:leibfriedb}.

A drawback of current microfabricated ion traps stems from the ions at the rf null naturally forming a 1-dimensional string, so far, limiting their usefulness for applications that require the formation of arbitrary 2-dimensional (2D) ion lattices. Penning traps offer a platform for a 2D lattice of ions, but the rotating crystal makes individual ion addressing and readout experimentally challenging, and the lattice geometry is limited to the naturally forming Wigner crystal \cite{Britton2Dsim}.

The applications for a 2D ion lattice are numerous and far reaching, including, among others, spatial magnetic and electric field sensing \cite{Kotler, narayanan:114909, 12:brownnutt}, force detection \cite{forcedetect}, interactions between neutral atoms and ions \cite{BECnIon} and cluster state quantum computing \cite{ClusterStateQC}.

A further exciting application for a 2D ion lattice is in the field of analogue quantum simulation where the Hamiltonian of a complicated many-body system can be realised and its properties measured \cite{schaetzRev,Nationtrap}. A promising approach to realising a 2D quantum simulator constitutes the proposal to use a two-dimensional lattice of individual rf ion traps \cite{CiracZoller,PhysRevLett.92.207901,PhysRevA.77.022324}. Each lattice site contains a single ion which interacts with neighbouring ions via the Coulomb force.
The current challenge lies in developing microfabrication techniques that allow for a tightly spaced ion lattice to be obtained where the ion-ion separation is small enough for coherent interactions to be measured. To date, experimental progress towards such a lattice has been limited to trapping dust particles and ion clouds above PCB boards \cite{2Ddust} or wire meshes \cite{clark:013114}. The optimum trap geometry of such a device has been well studied \cite{Sivernsa, PhysRevLett.102.233002}.

In operating a microfabricated ion surface trap, restrictions on the rf voltage that can be applied due to low flashover voltages exist \cite{NatGaaS,IQT2} and prohibit a large trap depth and ion-electrode distance, $d$. This limits the achievable ion lifetime and secular frequencies and results in large heating rates of the ion motion which scales as $d^{-4}$ \cite{PhysRevLett.97.103007} where heating has adverse effects on the fidelity of motion dependant qubit gate operations \cite{Sorensen}. The ability to apply large voltages to MEMS devices also has application in nanoelectrospray thruster arrays for spacecraft \cite{Paine2004112,krpoun,4443818,4801975}, where high electric fields are desirable and may also be useful in microfabricated mass-spectrometry \cite{Syms} as well as particle and cell sorting \cite{Grujic}.

Here, we report on a general microfabrication process to achieve large breakdown voltages in microfabricated devices and use this process to fabricate a 2D lattice of ion traps on a microchip. This process could also be used for other surface electrode ion trap geometries. Using this chip we demonstrate trapping of a 2D lattice of single Yb$^+$ ions, the deterministic trapping of multiple ions at lattice sites and rudimentary ion shuttling between lattice sites. The device presented is suitable for 2D sensing applications.
The ion-ion distance for this device is however too large to be used for quantum simulation. We discuss how a minor modification to the microchip design would allow for simulation of 2D spin lattices.
This modified design would also offer possible architectures for 2D cluster state generation  \cite{ClusterStateQC} and general quantum information processing \cite{CiracZoller}.  Since the device presented in this article is an integrated ion trap chip \cite{IQT2} it is potentially scalable to much larger ion lattices.

\begin{figure*}[t]
\centering
\includegraphics[scale=1]{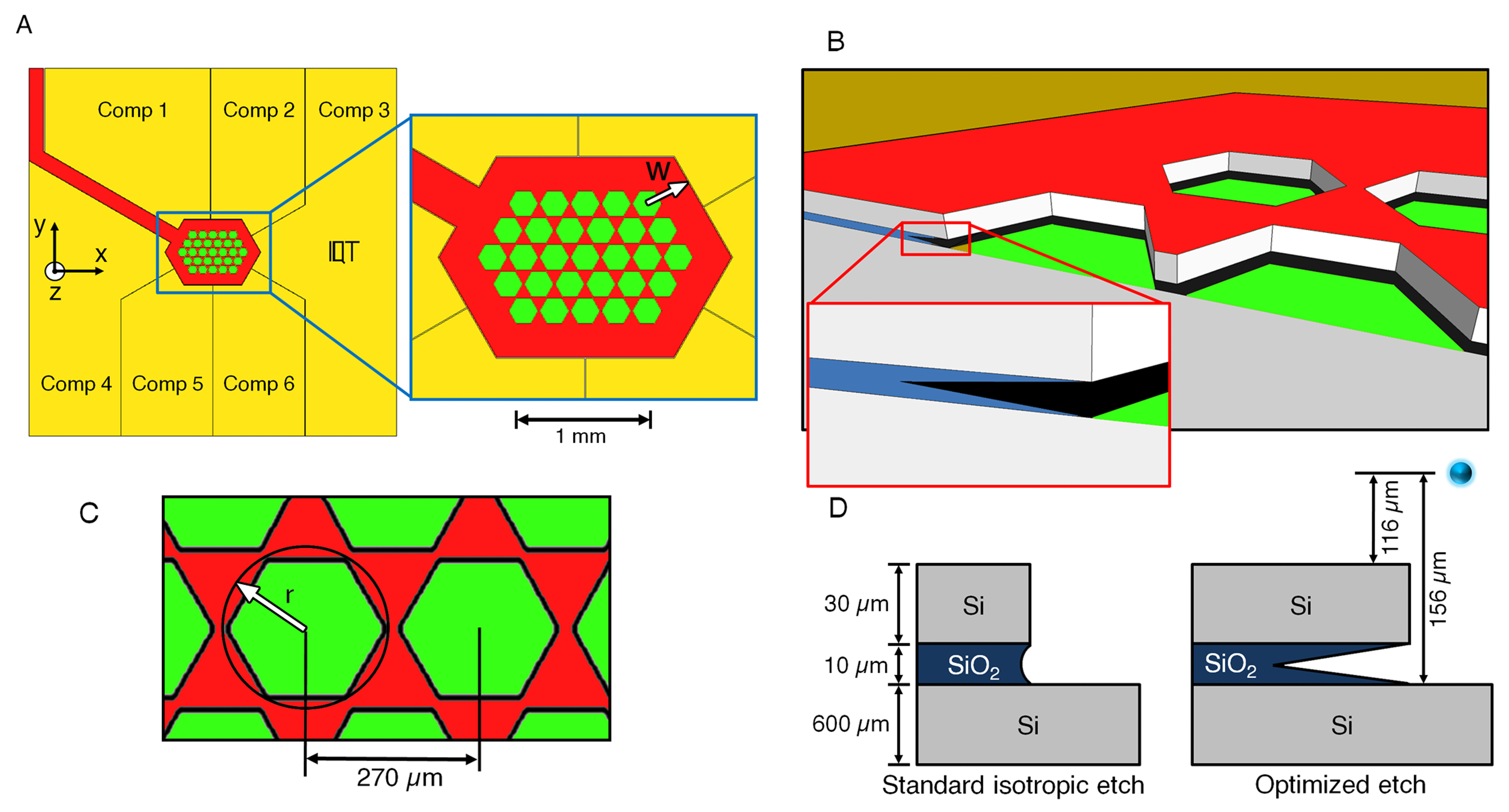}
\caption{\textbf{Design of the lattice microchip electrode geometry and electrode undercut} (\textbf{A}) The electrode design for the microchip. There are 29 trapping sites, one above the centre of each hexagon of ground electrode. The rf electrode is shown in red, compensation electrodes are shown in gold and the ground electrode is shown in green. The distance from the outer polygon centre to the rf edge, labeled w, is 410 $\mu$m. (\textbf{B}) Cross-section drawing of the trap geometry, showing the SOI layered geometry. The close-up inset shows the deep V undercut achievable with our fabrication process. (\textbf{C})  Close-up of the hexagonal trap geometry with a polygon radius r = 125 $\mu$m and neighbouring traps separated by 270.5 $\mu$m. (\textbf{D}) Two schematics of the etch profile, a typical isotropic etch is shown on the left, showing a slight undercut into the buried oxide. The right picture shows the profile of our optimised etch. The preferential etch rate along the bond face results in a deep V-shaped undercut. The ion height relative to the trap electrodes is also shown.}
\label{design}
\end{figure*}

\section{Results}

{\bf Fabrication of the microchip.} The layered structure of a commercial silicon-on-insulator (SOI) wafer was used by Britton \textit{et al.} \cite{britton:173102} to fabricate a linear ion trap permitting the application of voltages of up to 150 V. We have developed a different fabrication process using an SOI wafer to fabricate a planar ion trap with a significantly increased trap depth whilst maintaining large ion-electrode separations. This was achieved by a combination of recessed electrodes and the application of voltages in excess of 1 kV. We describe both of these fabrication advances in more detail below.

We utilise the layered SOI wafer to form a planar trap electrode structure with a 2D electrode pattern, shown in Fig. \ref{design}A. However, instead of the ground electrodes lying on the same plane as the rf electrodes they are recessed from the chip surface. This recess was achieved by using the SOI handle layer as rf ground, shown in Fig. \ref{design}B. By using recessed electrodes the trap depth for a given ion-electrode separation and voltage can be substantially increased. Each recess in the rf electrode corresponds to a single trap site and is hexagonal in shape, the recessed hexagonal dimensions are shown in Fig. \ref{design}C.

\begin{figure*}[t]
\centering
\includegraphics[scale=1]{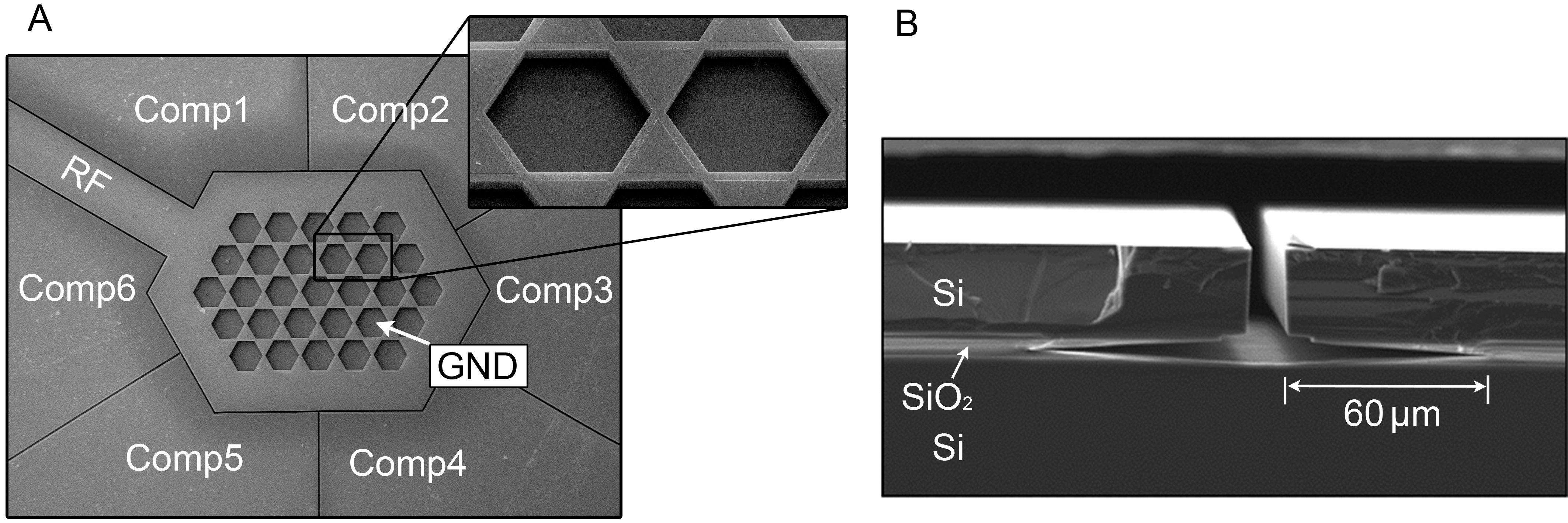}
\caption{\textbf{Scanning electron micrographs of the fabricated microchip} (\textbf{A}) An SEM image of a finished microchip. The inset shows a close up of two of the hexagonal traps, this shows the recessed ground electrode. (\textbf{B}) An SEM image of the cross section of the layered SOI structure at the interface between two compensation electrodes. The deep V-shaped undercut into the oxide layer is clearly visible extending 60 $\mu$m into the SiO$_2$ layer.}
\label{electrodes1}
\end{figure*}

To ensure ions can be trapped at a large ion height while maintaining a deep trapping potential using microfabricated ion traps in general and our lattice ion trap specifically a fabrication process that allows for high voltages to be applied was developed. This process not only has use for ion trap fabrication but for other microfabricated and microelectromechanical systems (MEMS) devices in general. We chose a thick oxide layer thus increasing the path length between the electrodes and the handle layer as well as reducing capacitance between the rf electrode and ground. However this would typically only result in a modest increase in breakdown voltage. We were able to obtain a much larger increase by using a specialised fabrication process. In order to explain this process it is important to note that voltage breakdown will usually occur via insulator surfaces connecting two conductors rather than through the insulator bulk. We selected a particular SOI structure that was fabricated by wafer bonding two wafers with 5 $\mu$m thick oxide surfaces to form  a 10 $\mu$m thick buried oxide layer (see Methods). Using a buffered HF etch, lasting $130-140$ minutes, the oxide layer was etched to expose the handle layer. As there is an increased etch rate on the interface where the two oxide layers were bonded together a highly anisotropic etch results laterally under the electrodes. Rather than obtaining the usual approximately straight etch profile, we were able to obtain the V-shaped undercut as shown in Fig. \ref{design}D. This V-shaped etch profile substantially increases the path length between the electrodes and ground resulting in extremely high breakdown voltages as it increases the effective distance for surface flashover from 10 $\mu$m to up to 120 $\mu$m. Due to the depth of the undercut, the hexagon structure becomes completely detached from the handle layer, supported by the thick surrounding rf edge. This greatly reduces the number of potential flashover initiation points, such as sharp points and defects, and aids the improvement in the flashover voltage. The finished trap structure can be seen in Fig. \ref{electrodes1}A, with the undercut shown in  Fig. \ref{electrodes1}B. Electrical flashover measurements were performed on SOI test samples to determine the voltage that can be applied (see Methods). The mean flashover voltages were measured to be $V_{\textrm{dc}}$ = 1300(80) V and $V_{\textrm{rf}}$ = 1061(10) V, where the error is the standard deviation. These values are one or two orders of magnitude higher than in previously fabricated microfabricated ion traps \cite{IQT2, britton:173102}. In fact these voltages correspond to typical values applied in traditional ion traps made using large metallic rods illustrating the impressive performance of our microchip.

\begin{figure*}[t]
\centering
\includegraphics[scale=1]{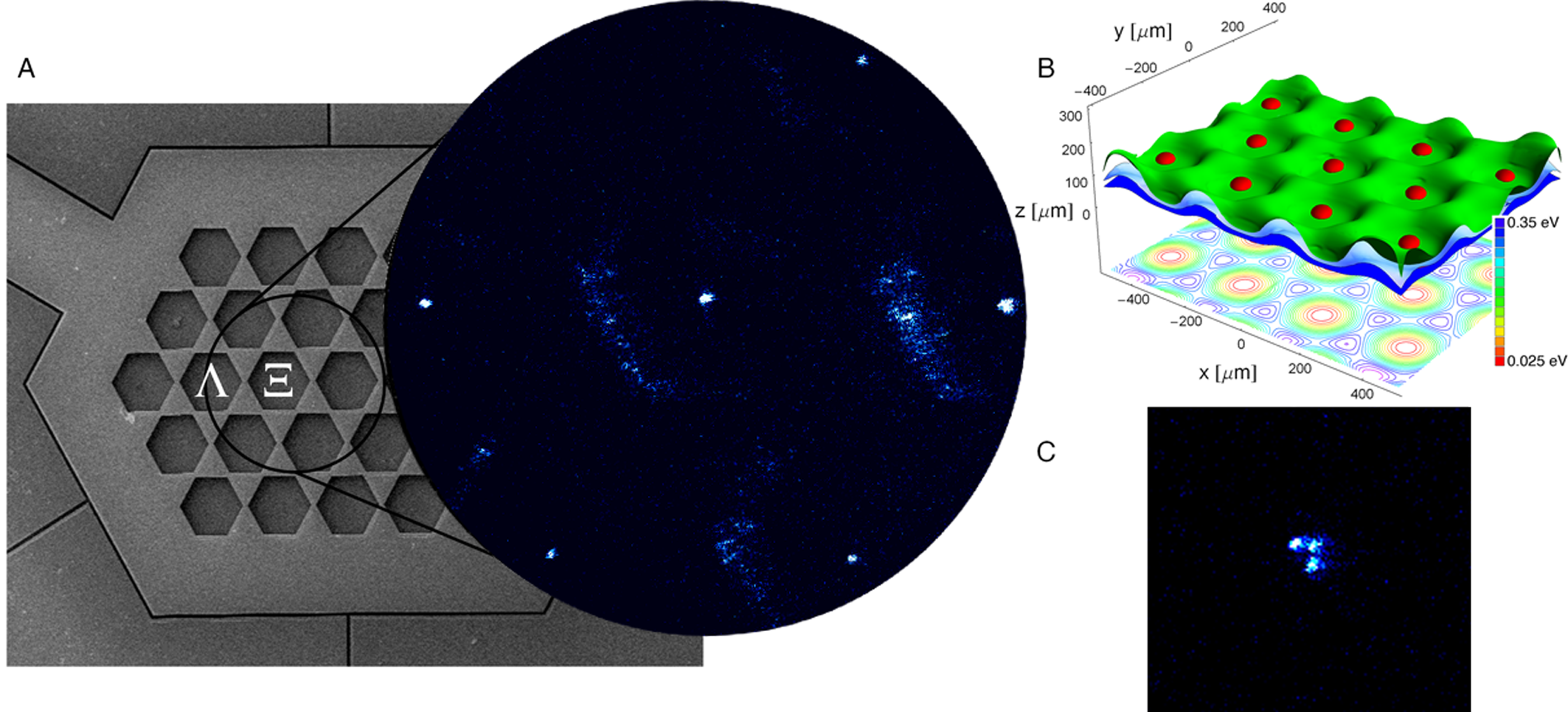}
\caption{\textbf{Image of trapped ion lattice on the microchip} (\textbf{A}) An image of six ions trapped simultaneously on the lattice is shown, the hexagonal shaped electrodes are also visible due to laser scatter. The area imaged was limited by our current imaging system and marked by a black circle over an SEM image of the trap. (\textbf{B}) A 3D contour plot of the lattice potential for the central 11 trap sites with an rf voltage of 455 V. The equipotential surfaces correspond to: red = 0.04 eV, green = 0.4 eV, light blue = 0.5 eV and blue = 0.6 eV. A 2D contour plot of the potential at the ion height is projected below, the equipotential lines are separated by 0.04 eV. (\textbf{C}) An image of a single lattice site containing a three-ion Coulomb crystal.}
\label{6ions}
\end{figure*}

The microchip consists of 29 traps arranged in a triangular lattice, with each ion having up to six nearest neighbours with an ion-ion separation of 270.5 $\mu$m. The ion-electrode separation is 156 $\mu$m. Due to the recessed ground electrodes the ion height from the top of the trap surface is 116 $\mu$m, as seen in Fig. \ref{design}D. A 3D contour plot of the trap potential relating to the chip in Fig. \ref{6ions}A, for the central 11 trap sites, is shown in Fig. \ref{6ions}B. A 2D contour plot is projected below showing the potential at the ion height. To adjust the position of the ions to compensate for stray electric fields and perform shuttling operations quasi-static voltages can be applied to six electrodes (Comp 1-6) which surround the rf electrode.

{\bf Trapping and manipulation of ions on the microchip.} An rf voltage $V_{\textrm{0}}$ = 455(3) V at a frequency $\Omega/2\pi$ = 32.2 MHz was applied to the rf electrode to produce a trapping potential. All other electrodes were initially held at ground. A flux of neutral ytterbium atoms travelling parallel to the trap surface was produced by ohmically heating a natural abundance ytterbium oven from which $^{174}$Yb atoms were resonantly ionised using a two-colour photoionisation process with light at 399 nm and 369 nm. Trapped ions were then Doppler cooled by the 369 nm light, with 935 nm light used to repump the ion from a long-lived D-state \cite{IQT1}. The cooling and ionisation laser beams were highly elliptical, forming a light sheet co-propagating parallel to the trap surface. An image plane co-incident with the plane of trap centres above the microchip was imaged onto an electron multiplying CCD array to view the ions.

The ion secular frequencies along the trap principle axes were measured to be  $(\omega_{x'},\omega_{y'},\omega_{z'})/2\pi$ = $(1.58, 1.47, 3.30) \pm 0.01$ MHz respectively, for a single ion trapped in the top right corner of the array. Using these measurements, numerical simulations of the trap predict a trap depth of 0.42(2) eV. The lifetime of a single laser cooled ion was $\sim$ 90 minutes, likely limited by background collisions and $\gtrsim$ 5 minutes without cooling light.

A 2D ion lattice can be seen in Fig. \ref{6ions}A, six ions are trapped in adjacent lattice sites, with the viewable area limited by our imaging system. Due to the stochastic nature of ion loading, trapping a uniform lattice is difficult when uniformly illuminating all trap sites. However by steering the 399 nm ionisation laser to address individual sites we can controllable fill the lattice.  We can also introduce defects, either empty sites or multiple ions per site, in the same manner. A three-ion defect is shown in Fig. \ref{6ions}C. Such lattice defects may be of interest when investigating Bose-Hubbard physics, such as superfluid-Mott insulator transitions or simulations of spin models with spin greater than 1/2.

\begin{figure*}[t]
\centering
\includegraphics[scale=1]{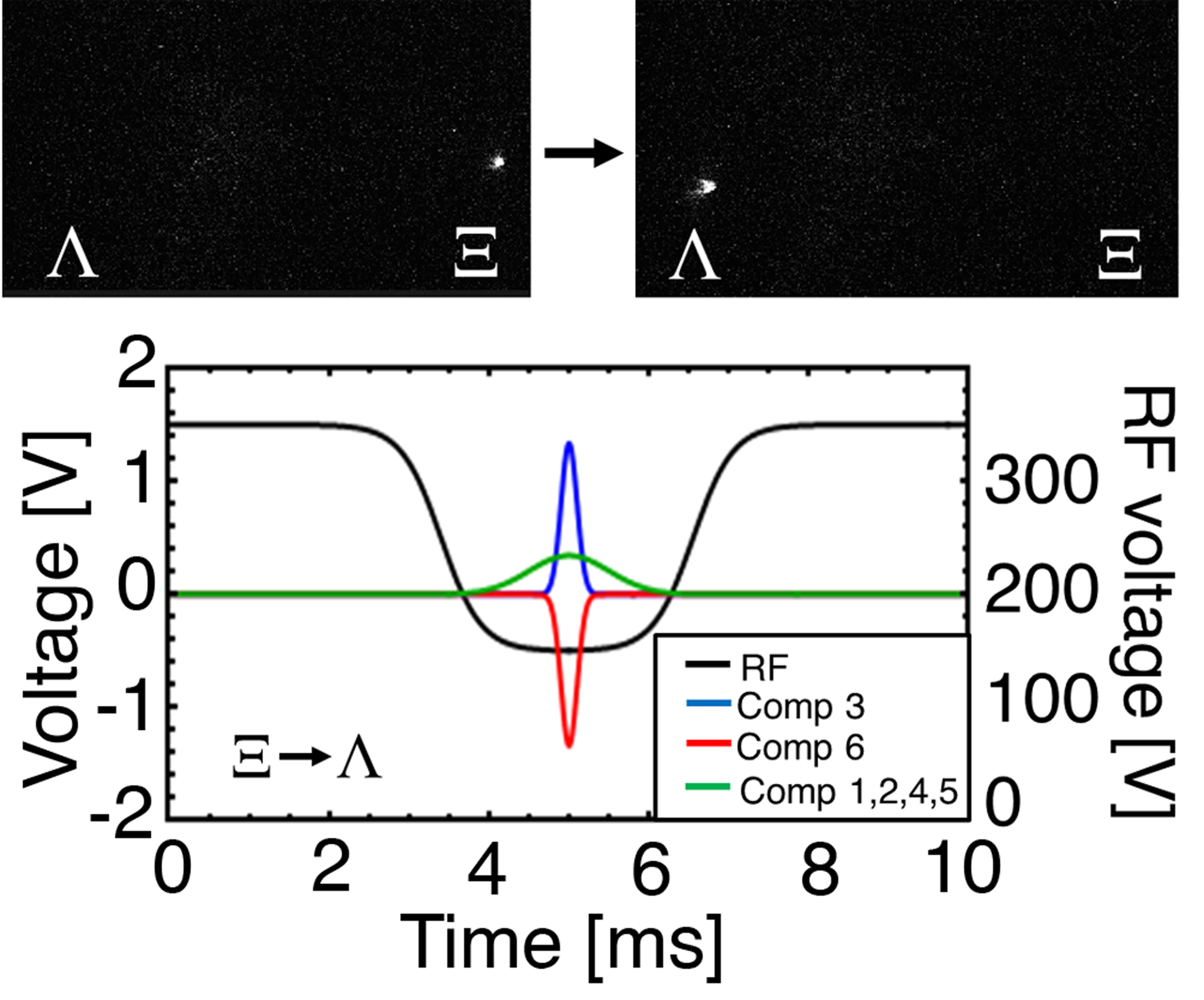}
\caption{\textbf{Ion shuttling on the microchip} Two images are shown before and after shuttling an ion from lattice site $\Xi$ to $\Lambda$. The shuttling voltage profile for this shuttling operation is shown. To shuttle the ion back to $\Xi$ the polarity of the voltages on Comp 3 and 6 is reversed.}
\label{shuttling}
\end{figure*}

We have also demonstrated shuttling of ions between different sites of the lattice. An ion is initially trapped at site $\Xi$, as marked in Fig. \ref{6ions}A. By lowering the rf voltage to minimise the potential barrier and applying control voltages to the surrounding electrodes, the ion is shuttled to position $\Lambda$. Figure \ref{shuttling} shows two images of the ion before and after shuttling, also shown are the typical shuttling voltages. To shuttle the ion back to $\Xi$ the polarity of the voltage on Comp 3 and Comp 6 is reversed. While this shuttling was performed using global control electrodes, enhancing the trap design by adding local control electrodes to the handle layer around each trapping site using backside etches and depositions would allow the distribution of ions over the lattice sites to be changed after the ions are loaded. This would enable separate loading and experiment zones, and for the lattice to be repaired if an ion is lost due to a background collision. Any heating of the ions due to such shuttling can be removed by laser cooling prior to temperature sensitive operations. Such local control would also enable precise control of the micromotion at each site.

\section{Discussion}

A 2D lattice such as we demonstrate here has many potential future applications. The use of single ions for highly sensitive magnetic-field sensing has been demonstrated \cite{Kotler}, and the use of a lattice would allow for spatial as well as temporal measurements \cite{12:brownnutt}. Single ions have been proposed to be used as point imperfections inside Bose-Einstein condensates \cite{BECnIon}, and ion lattices would allow the effect of precisely tailored arrays of imperfections, both regular and irregular depending on microchip design, to be studied.

One important future application for a 2D ion lattice is in the field of quantum simulation, where trapped ions represent spins in solid-state systems. Interactions between spins can be tailored to create analogues of different 2D systems (such as Ising or Heisenberg spin lattices) in a clean and controllable manner, whose properties can be determined by measurement of the state of all the ions in the array using a light sheet and a CCD array, something that is difficult or impossible to perform in actual solid-state systems.

In simulating these systems, interactions between ions trapped in adjacent sites $a$ and $b$ would be mediated by the Coulomb interaction. The coupling between motional states $H_{\rm Coul}=-\hbar\Omega_{\rm ex}/2(a^\dagger b+ab^\dagger)$ where $a,a^\dagger,b,b^\dagger$ are the phonon annihilation and creation operators for the two sites, and the coupling strength
\begin{equation}
\Omega_{\rm ex} = \frac{e^2}{2\pi \varepsilon_0 m\omega}\frac{1}{s^3}
\label{omex}
\end{equation}
for a pair of singly-charged ions of mass $m$ confined in traps with secular frequency $\omega$ separated by distance $s$ \cite{BrownNIST,Harlanderantenna}.

For the lattice we have presented here, where $s=270.5\,\mu$m and with a trap frequency of 1$\,$MHz, the coupling strength $\Omega_{\rm ex}/2\pi$ would be only 2 Hz, impractical for performing simulations and very small compared to the strengths already experimentally demonstrated using linear surface traps \cite{BrownNIST,Harlanderantenna}. The coupling strength has a very strong dependence on the ion-ion separation ($\Omega_{ex}\propto s^{-3}$) and reducing this distance would be key to obtaining useful couplings. A lattice based on smaller ion-ion distances can easily be designed and produced using the fabrication method presented in this paper.

Using an SOI wafer with a 2.5$\,\mu$m device layer and a 2.5$\,\mu$m oxide layer fabricated by wafer bonding two 1.25$\,\mu$m oxide layers together would reduce the depth of the ground plane beneath the surface of the microchip to 5$\,\mu$m, and a trap electrode design with a hexagon radius of 13$\,\mu$m and hexagon centre separation of 32$\,\mu$m fabricated on such a wafer would produce a lattice of traps 15$\,\mu$m above the surface of the microchip. To reduce scatter from tightly focussed laser beams, an etch is performed outside the trapping area to lower the surface by 100 $\mu$m. For a trap frequency $\omega/2\pi$ of 1 MHz, the motional coupling strength $\Omega_{\rm ex}/2\pi$ would then be 1.3 kHz, almost a thousandfold increase compared to our current chip design. This value is comparable to previously achieved coupling strengths between individual wells in linear traps \cite{BrownNIST,Harlanderantenna}.

If this trap were used to simulate a 2D spin lattice using the scheme proposed by Porras and Cirac \cite{PhysRevLett.92.207901, QSerror} a spin-spin coupling strength $J\approx2\pi\times1.2$ kHz would be obtainable using a pulsed 355 nm laser with an average power of 920 mW (see Methods). It is also important to consider possible sources of decoherence to evaluate if a successful quantum simulation is expected for this prospective trap. We consider both heating of the motion of the trapped ions and photon scattering due to the laser driving the gate. These rates should be low compared with the spin-spin coupling strength $J$. At such small ion-electrode distance, heating rates at room temperature will be many orders of magnitude greater than $J$, necessitating cooling of the chip. While estimating heating rates is imprecise, based on previous measurements in gold traps \cite{Labaziewicz08} an estimate for a trap of this design operating cryogenically at 7 K is $\dot{n}$=180 s$^{-1}$ (see Methods). The photon scattering due to the coupling laser is estimated to be 27$\,$s$^{-1}$. Both these rates are significantly less than $J$ indicating that this type of quantum simulation should be possible with this combination of chip and driving laser.

We have demonstrated a 2D ion lattice integrated on a microchip along with a fabrication process to significantly increase the voltage that can be applied to microfabricated devices. By increasing the breakdown voltage, both 2D lattice and linear surface traps with higher trap depth and lower heating rates (by making use of larger ion-electrode distances) can be fabricated. For surface traps built to perform quantum information processing this reduces the effects of motional heating on gate fidelities \cite{Sorensen}. Additionally, we have used this advance to demonstrate the operation of a 2D lattice of ytterbium ions on a chip which offers many potential uses including spatially resolved force measurement in two dimensions and magnetic and electric field sensing, and with modifications may offer a platform for quantum simulations.

\section{Methods}
{\bf Wafer fabrication.} The SOI wafers are produced by Ultrasil Corporation who use room temperature fusion bonding to bond the oxide surfaces before the substrate is annealed at 980$^\circ$C. X-ray reflection studies of bonded SOI wafers show that even after annealing several monolayers of water molecules remain at the bond-interface in the structure of the SiO$_2$ \cite{01:Rieutord}. This results in the V-shaped undercut caused by the anisotropic preferential etch along this interface.

{\bf Surface flashover measurements.} Surface flashover measurements were performed on test chips consisting of gold coated silicon islands surrounded by the exposed handle layer. The chips were glued, using conductive silver glue, onto a ceramic chip carrier and placed in high vacuum at a pressure of 6 $\times$ 10$^{-4}$ Pa. Radio-frequency voltage was applied by connecting a quarter-wave helical coil resonator to the vacuum feedthrough. This ensured $>$ 90\% coupling between the rf amplifier and chip at a resonant frequency of 28.0 $\pm$ 0.5 MHz with $Q$ = 210 $\pm$ 15. The rf voltage was steadily increased while observing the sample. Upon breakdown the voltage was measured using a capacitively coupled probe.

Radio-frequency flashover was observed to occur $\approx$ 20\% lower than static flashover. The exact cause for this is not known, however it is likely a result of local heating on the chip and outgassing from the surface when the rf voltage is applied. To apply sufficient voltage to trigger flashover, up to 30 W of rf power was applied to the chip. This power would be predominately dissipated on chip, which has the largest impedance. This would result in significant heating of the chip and the release of gases previously adsorbed to the surface.

{\bf UHV microchip mounting and high voltage application.} The trap was mounted onto a ceramic chip carrier, with gold ribbon wire connecting the electrodes to the chip carrier bond pads. The chip capacitance was measured to be 19 $\pm$ 1 pF. The chip was mounted inside a UHV vacuum system at a pressure of 8.0$\times$10$^{-8}$ Pa. To ensure good rf grounding 820 pF capacitors were wire bonded to the static electrodes. The radio frequency voltage was applied via an external helical coil resonator with Q = 160 to provide voltage amplification and filtering of unwanted frequencies \cite{Sivernsb}.

{\bf Quantum simulation parameters for modified microchip design.} The Porras and Cirac spin simulation scheme \cite{PhysRevLett.92.207901, QSerror} requires a state dependant force to be applied to the ions whose internal states encode the simulated spins. The effective spin-spin coupling rate is given by $J=\beta F^2/4\hbar m_i\omega_i^2$ where $\hbar=h/2\pi$, $h$ is Planck's constant, $\beta=\Omega_{ex}/\omega_i$ and $F$ is the magnitude of the state dependant force applied to each ion \cite{PhysRevLett.92.207901}.
For the proposed modified chip design, with its ion-ion separation of 32 $\mu$m and using a pulsed laser with a wavelength of 355 nm divided into counterpropagating beams of total average power of 920 mW focused into light sheets of width 50 $\mu$m and depth 15 $\mu$m to address a $3\times3$ lattice of ions, a coupling strength $J\approx2\pi\times1.2$ kHz could be achieved. Since the simulation is driven by the displacement of the ions due to the force $F$, the simulated spins are entangled with the motional state of the ions at the end of the simulation. This produces an error in the measured spin properties of the simulated system proportional to $F^2$ \cite{QSerror}. For the proposed modified design we calculate a simulation error due to this effect of $\approx$ 25 \% for the measurement of single-particle properties of the simulated system. In addition to using lasers to produce the spin-spin coupling, magnetic field gradients could also be used to generate the required spin-dependent forces \cite{MagfieldGrad} and would allow individual addressing of ions.

Motional heating of an ion is a result of electric field fluctuations at the trap secular frequency, and depends strongly on the ion-electrode distance, which in this case is 19$\,\mu$m~(the ion lies 15$\,\mu$m~above the surface of the chip, however the electrode under the ion is recessed by 5$\,\mu$m; the closest points to the ion lie on the edges of the hexagonal cut-out). For the proposed modified chip we assume that the electric-field noise density is the same as measured previously for a gold-electroplated trap at a temperature of 7 K \cite{Labaziewicz08}, after adjusting for the different ion-electrode distance $d$  and trap frequency $\omega$ using a scaling of $d^{-4}\omega^{-1}$. For our ion-electrode distance of 19$\,\mu$m and trap frequency $2\pi\times 1\,$MHz we estimate an electric-field noise density $S_E(\omega)\approx 5.3 \times10^{-12}$ V$^2$m$^{-2}$Hz$^{-1}$ .  This corresponds to a heating rate $\dot{n}=180\,$s$^{-1}$ for an $^{171}{\rm Yb}^+$ ion.

\section{Acknowledgments}
We would like to acknowledge helpful discussions with D. Porras and J. D. Siverns. This work is supported by the UK Engineering and Physical Sciences Research Council (EP/E011136/1, EP/G007276/1), European Commission's Seventh Framework Programme (FP7/2007-2013) under grant agreement no. 270843 (iQIT), the European Commission's Sixth Framework Marie Curie International Reintegration Programme (MIRG-CT-2007-046432), the Nuffield Foundation and the University of Sussex.

\section{Author contributions}
R.C.S. planned the experiment, designed the microchip and the fabrication process, performed the experiment and wrote the manuscript, H.R. fabricated the microchip, S.W., K.L. and S.C.W. performed the experiment and assisted in writing the manuscript, P.S. and M.K. assisted in the development of the fabrication process and W.K.H. planned the experiment, coordinated the work, helped design the microchip and fabrication process and assisted in writing the manuscript.

\section{Competing Financial Interests}
The authors declare that they have no competing financial interests.

\end{document}